\documentclass[sigconf]{acmart}

\usepackage{booktabs}
\usepackage{multirow}
\usepackage{amsmath,amsfonts}

\usepackage{amssymb}
\usepackage{graphicx}
\usepackage{textcomp}
\usepackage{xcolor}
\usepackage{microtype}
\usepackage{xspace}
\usepackage{enumitem}
\usepackage{tikz}
\usepackage{afterpage}
\usetikzlibrary{shapes,arrows,positioning,fit,backgrounds}

\newcommand{\DIG}{\textsc{DIG}\xspace}

\newcommand{\eu}{\mathbf{e}_u}
\newcommand{\ev}{\mathbf{e}_v}
\newcommand{\esid}{\mathbf{e}_{\mathrm{sid}}}
\newcommand{\cuv}{\mathbf{c}_{u,v}}
\newcommand{\utot}{\mathbf{u2t}}         
\newcommand{\utothat}{\hat{\mathbf{u2t}}} 
\newcommand{\utotbar}{\bar{\mathbf{u2t}}} 
\newcommand{\lrank}{\mathcal{L}_{\mathrm{rank}}}
\newcommand{\lrecall}{\mathcal{L}_{\mathrm{recall}}}
\newcommand{\lcommit}{\mathcal{L}_{\mathrm{commit}}}
\newcommand{\lsem}{\mathcal{L}_{\mathrm{sem}}}
\newcommand{\lut}{\mathcal{L}_{\mathrm{u2t}}}

\begin{document}

\title{Discrimination Is Generation: Unifying Ranking and Retrieval from a Tokenizer Perspective}

\author{Shuli Wang}
\email{shuliw1996@gmail.com}
\renewcommand{\shortauthors}{Wang et al.}

\begin{abstract}
Semantic IDs (SIDs) define the generation space of generative recommendation and directly determine its personalization ceiling. However, existing tokenizers are trained independently with retrieval objectives, leaving personalization signals fully decoupled from the SID construction process---a fundamental gap that causes generative retrieval to persistently lag behind discriminative ranking. In this paper, we rethink the essence of SIDs: \emph{ranking seeks argmax in item space while retrieval seeks argmax in token space; both are the same problem solved at different granularities.} Based on this insight, we propose \DIG (\textbf{D}iscrimination \textbf{I}s \textbf{G}eneration), which embeds the tokenizer inside a discriminative ranking model for end-to-end training---the ranker naturally becomes a retrieval model, yielding two models from a single training run. \DIG is organized around a \emph{feature assignment taxonomy}: item-intrinsic static features are encoded into SIDs, user-item cross features (u2i) implicitly drive codebook boundaries toward recommendation decision boundaries during training, and an MLP$_\mathrm{u2t}$ distillation module approximates u2i at the token level for inference. Experiments on three public benchmarks and two industrial datasets demonstrate that \DIG simultaneously improves ranking, retrieval, and unified retrieval-ranking quality.
\end{abstract}



\maketitle

\section{Introduction}
\label{sec:intro}

Generative recommendation (GR) quantizes items into discrete Semantic ID (SID) sequences and performs full-corpus argmax via beam search---complexity decoupled from the item catalog size, fundamentally breaking the computational bottleneck of traditional retrieval funnels~\cite{rajput2023tiger,li2024letter}.
Building on the SID$+$NTP framework pioneered by TIGER~\cite{rajput2023tiger}, subsequent work has converged on a multi-stage pipeline: the tokenizer is trained independently, and the generative model is aligned downstream.
Quality improvements have come through collaborative signal alignment~\cite{li2024letter,zheng2024das,yin2026dos,du2024etegrec}, end-to-end SID learning~\cite{du2024etegrec,lin2024resid}, and differentiable quantization~\cite{fu2026differentiable}.

Despite this progress, a critical deficiency persists: \textbf{existing SIDs are severely under-personalized.}
Current SID methods fall into two categories: \emph{pure semantic tokenization} (e.g., TIGER---pretrained language model + RQ-VAE quantization) and \emph{retrieval-aligned tokenization} (e.g., LETTER, DAS, DOS---dual-tower contrastive losses driving quantization).
Both encode only static item attributes; user-item cross features ($\mathbf{c}_{u,v}$)---the very signals that give discriminative rankers their fine-grained personalization power---\emph{never participate in codebook construction.}
The consequence: the same item receives an identical SID regardless of whether it faces a highly matched or completely mismatched user.
These personalization signals are precisely the core source of discriminative ranking ability---a structural ceiling that has kept generative retrieval persistently trailing discriminative ranking.

The root cause is architectural: \textbf{tokenizers are trained with retrieval objectives and fully decoupled from discriminative signals.}
Retrieval objectives only require that related items be close in embedding space; they impose no fine-grained personalization distinctions.
The two-stage pipeline ensures that discriminative gradients can \emph{never} flow back into the codebook, and personalization signals can \emph{never} participate in quantization.
Two recent directions attempt to close the gap---aligning tokenizers with downstream generation objectives~\cite{du2024etegrec,lin2024resid} and incorporating semantic tokens into rankers~\cite{chen2025store}---but both still optimize within the retrieval paradigm and neither injects discriminative ranking gradients back into SID construction.
Both paths converge on a more fundamental question: \textbf{can the objective of discriminative ranking directly drive SID codebook construction, enabling the same token set to serve both generative retrieval and discriminative ranking?}

We answer this by rethinking the essence of SIDs.
Both tasks solve the same optimization: ranking finds the highest-scoring item in \emph{item space} while retrieval finds the optimal path in \emph{token space.}
The SID sits exactly at the intersection of both spaces---not merely a preprocessing step for generative recommendation, but a \textbf{bridge between discriminative and generative recommendation}.
If SID construction is driven directly by a discriminative objective, the same token representation can simultaneously carry ranking discriminability and retrieval generativity.
In other words, \textbf{generative retrieval capability does not require a separate generative model---it is already latent inside the discriminative ranker, and the tokenizer's role is to release it.}

\begin{figure}[t]
  \centering
  \includegraphics[width=\linewidth]{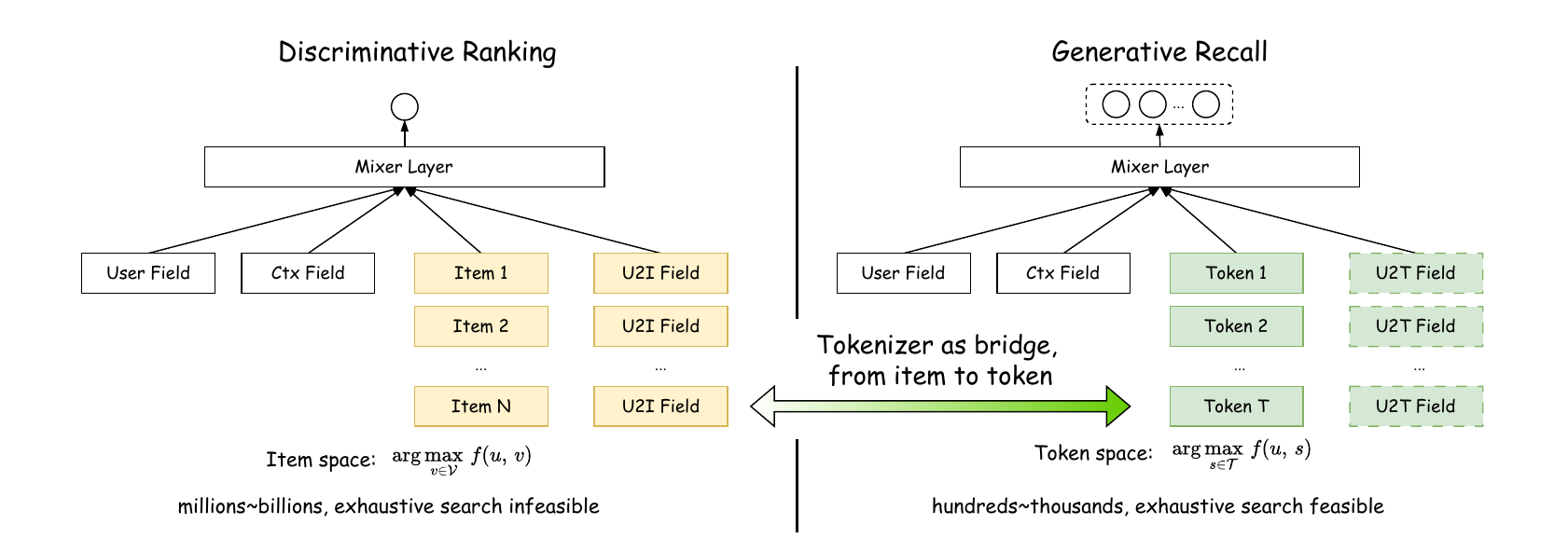}
  \caption{Comparison of existing generative retrieval pipelines vs.\ \DIG. Existing tokenizers are trained with retrieval objectives (reconstruction or contrastive loss), fully decoupled from discriminative ranking signals. \DIG embeds the tokenizer inside the ranker for end-to-end joint training, releasing generative retrieval capability already latent in the discriminative model.}
  \label{fig:motivation}
\end{figure}

Based on this insight, we propose \textbf{\DIG} (\textbf{D}iscrimination \textbf{I}s \textbf{G}eneration).
\DIG embeds the tokenizer inside a discriminative ranking model for end-to-end joint training---the ranker naturally becomes a retrieval model, yielding two models from a single training run.

To make this bridge structurally sound, \DIG is organized around a \textbf{feature assignment taxonomy} as its central axis, systematically resolving how three categories of features are handled across training and inference.
\textbf{Unified Tokenizer} (\S\ref{sec:tokenizer}): the VQ encoder takes only item-side static features; an offline Balanced K-Means tree guarantees zero-collision stability across model versions; SIDs are decoupled from SID embeddings---SIDs handle addressing while SID embeddings carry semantic expression end-to-end, resolving the fundamental trade-off between discrimination and semantic sharing.
\textbf{Unified Training} (\S\ref{sec:training}): ranking and retrieval share a single parameter set driven by a five-part unified loss; u2i cross features are concatenated into the scoring MLP and implicitly drive codebook boundaries toward recommendation decision boundaries end-to-end; MLP$_\mathrm{u2t}$ uses the ranker's u2i modeling capacity as teacher signal to learn user-conditioned u2t representations, substantially improving personalization resolution on the retrieval path compared to statistical averages.
\textbf{Unified Inference} (\S\ref{sec:inference}): MLP$_\mathrm{u2t}$ generates u2t online to recover the third feature type; the ranking model directly executes beam search for retrieval---the unified paradigm follows naturally.

\noindent\textbf{Contributions.} We summarize our contributions as follows:
\begin{itemize}[leftmargin=*,nosep]
  \item We rethink the essence of SIDs and propose a novel \emph{unified} paradigm: the tokenizer bridges item space and token space, and discriminative ranking already contains generative retrieval capability---no extra model is needed; the tokenizer releases it.
  \item We present \DIG, a systematic implementation centered on the feature assignment taxonomy, integrating tokenizer design, unified training, and inference alignment to inject discriminative signals end-to-end into SID construction, with MLP$_\mathrm{u2t}$ personalized distillation bridging the feature gap between ranking and retrieval.
  \item We conduct extensive experiments on five datasets (three public + two industrial). \DIG consistently outperforms state-of-the-art generative retrieval baselines on recall metrics and simultaneously improves ranking AUC, with particularly strong gains in sparse and u2i-rich scenarios.
\end{itemize}

\section{Related Work}
\label{sec:related}

\subsection{Generative Recommendation and Semantic IDs}

Generative recommendation quantizes items into discrete token sequences and models user preferences auto-regressively.
TIGER~\cite{rajput2023tiger} established the foundational SID$+$NTP two-stage framework: offline, Sentence-T5 embeddings are quantized via RQ-VAE~\cite{zeghidour2021soundstream} into SIDs; online, a Transformer auto-regressively generates the target item's token sequence.
Subsequent works improve SID quality primarily by introducing collaborative signals into the quantization process.
LETTER~\cite{li2024letter} aligns collaborative vectors with quantized semantic vectors through contrastive learning, producing more uniform codebook distributions.
CoST~\cite{yang2024cost}, DAS~\cite{zheng2024das}, and DOS~\cite{yin2026dos} adopt user-item dual-tower structures driven by collaborative labels as auxiliary supervision.
CoFiRec~\cite{wang2024cofirec} fuses semantic SIDs with collaborative IDs into coarse-to-fine hierarchical token representations.
All share a fundamental limitation: training signals come from retrieval-side contrastive learning, and the quantization process remains entirely decoupled from discriminative ranking objectives---u2i cross features $\mathbf{c}_{u,v}$ never participate in codebook construction.

ETEGRec~\cite{du2024etegrec} takes a step further by jointly training the tokenizer and generative model in an alternating frozen fashion, aligning them via NTP loss.
ReSID~\cite{lin2024resid} redesigns the tokenizer from an information-theoretic perspective, explicitly critiquing ETEGRec's \emph{self-referential problem}---SIDs are simultaneously training targets and model inputs, causing optimization instability when NTP loss adjusts SIDs end-to-end.
It is important to note that ReSID's critique targets ``using NTP loss to adjust SIDs''; \DIG's gradient comes from a discriminative BCE loss that predicts user clicks, not SID sequences, so the self-referential problem does not apply.
MTGRec~\cite{du2024mtgrec} enhances generative model robustness via multi-tokenizer data augmentation.
Differentiable SID~\cite{fu2026differentiable} connects SID tasks to NTP via Gumbel-Softmax, allowing NTP gradients to flow back into SID assignment.
Both adjust SIDs from a \emph{downstream NTP perspective}; \DIG reshapes the codebook from an \emph{upstream discriminative perspective}---the directions are opposite and complementary.

\subsection{Unifying Retrieval and Ranking}

Unifying retrieval and ranking into a single system is a long-standing research direction, and recent work has approached it primarily through generative models.
HSTU~\cite{zhai2024hstu} restructures recommendation as a sequence transduction task at trillion-parameter scale, verifying scaling laws for generative recommendation.
OneRec~\cite{qiu2024onerec} directly generates entire recommendation sessions using an encoder-decoder with MoE, replacing the traditional multi-stage cascade pipeline with a single generative system.
SynerGen~\cite{chen2024synergen}, UniRec~\cite{wang2026unirec}, OneRanker~\cite{sun2026oneranker}, and UniGRF~\cite{zhang2025ugrf} each explore how generative models carry both retrieval and ranking simultaneously, via multi-task losses, hierarchical decoding, or DPO post-training.
All \emph{start from a generative model} and embed ranking capability within it.
OneMall~\cite{jia2024onemall} takes a different approach, bridging two independent models via reinforcement learning where the ranker provides reward signals to the generative retriever.

\textbf{Fundamental distinction of \DIG.}
All of the above start from generative models to build unified frameworks; ranking capability is embedded into generation.
\DIG takes the \emph{opposite} direction: starting from a discriminative ranker and extending its scoring space to full-corpus retrieval through the tokenizer.
The ranker structure is unchanged; the tokenizer---acting as a bridge between item space and token space---endows the ranker with native retrieval capability.
This path fully preserves the discriminative ranker's accumulated advantages (u2i cross features, high-order feature interactions), achieves architecture simplicity through a single training run, and requires neither a generative paradigm nor RL bridging.
Generative capability is not designed in---it is \emph{released} from the discriminative ranker.

\subsection{Semantic Tokenization for Ranking}

The idea of semantic tokenization has recently migrated from generative retrieval into discriminative ranking, motivated primarily by storage and compute efficiency.
UIST~\cite{zhang2024uist} simultaneously tokenizes items and users, achieving $\sim$200$\times$ storage compression via hierarchical mixed inference; its tokenizer is trained independently with reconstruction error, decoupled from the ranking objective.
STORE~\cite{chen2025store} proposes Semantic Tokenization + orthogonal rotation + Efficient Attention, rotating low-cardinality static features into the SID embedding space to enhance feature interactions, achieving +2.71\% online CTR.
TRM~\cite{li2026trm} replaces item ID embeddings with semantic tokens in search and ranking at scale, reducing sparse storage by 33\% and achieving +0.85\% AUC.
These works demonstrate that semantic tokens can replace traditional item ID embeddings without hurting ranking quality---an important empirical foundation.
However, they share the same structural limitation as retrieval-side SID methods: the tokenizer is an independent preprocessing tool, u2i cross features $\mathbf{c}_{u,v}$ never participate in tokenizer construction, and the codebook boundaries reflect content similarity rather than recommendation decision boundaries.

\DIG differs from all of the above in three fundamental ways:
(1)~\textbf{Training signal}: discriminative BCE loss drives the tokenizer end-to-end, rather than reconstruction error or contrastive loss;
(2)~\textbf{Role of u2i features}: $\mathbf{c}_{u,v}$ implicitly drives codebook boundaries toward recommendation decision boundaries during training, making SID partitions reflect user preference rather than semantic similarity contours;
(3)~\textbf{System architecture}: \DIG proposes the unified paradigm where online retrieval directly reuses the ranking model via beam search, fundamentally eliminating the semantic gap of traditional dual-system pipelines.

\section{Methodology}
\label{sec:method}

\begin{table*}[!t]
\caption{Feature assignment taxonomy in \DIG. $\mathbf{x}_v^s$: item-side features; $\esid^{(1:l)}$: SID embedding prefix; $\mathbf{u2i}\triangleq\cuv$: item-level cross features; $\mathbf{u2t}^{(l)}$: token-level aggregation of $\cuv$ within bucket $s_l$.}
\label{tab:feature}
\centering
\setlength{\tabcolsep}{8pt}
\renewcommand{\arraystretch}{1.3}
\begin{tabular}{@{}llcll@{}}
\toprule
\textbf{Type} & \textbf{Symbols} & \textbf{In SID?} & \textbf{Training-side handling} & \textbf{Inference-side handling} \\
\midrule
Type-I: item-side        & $\mathbf{x}_v^s$              & \checkmark & VQ encoder input $\to\mathbf{e}_v\to\esid^{(1:L)}$         & Offline SID lookup table \\
Type-II: request-level   & $\mathbf{x}_u,\,\mathbf{ctx}$ & $\times$   & Ranking loss condition                                      & Beam search decoder condition \\
Type-III: struct.\ lossy & $\mathbf{c}_{u,v}$            & $\times$   & Implicit codebook shaping via batch $\utot^{(1:l)}$         & MLP$_\text{u2t}$ distillation $\to\utothat^{(1:l)}$ \\
\bottomrule
\end{tabular}
\end{table*}

We first present the problem formulation and \DIG's overall framework (\S\ref{sec:framework}), establishing the feature assignment taxonomy as the central design axis.
We then detail the Unified Tokenizer (\S\ref{sec:tokenizer}), Unified Training (\S\ref{sec:training}), and Unified Inference (\S\ref{sec:inference}).

\subsection{Problem Formulation and DIG Framework}
\label{sec:framework}

\noindent\textbf{Notation.}
Let $\mathcal{V}$ be the item set and $\mathcal{U}$ the user set.
Each item $v\in\mathcal{V}$ has static content features $\mathbf{x}_v^s$ (category, brand, region, etc.).
Each user $u\in\mathcal{U}$ has user features $\mathbf{x}_u$ (profile, behavior history) and context features $\mathbf{ctx}$ (time, location, scene).
For a user-item pair $(u,v)$, \textbf{u2i} cross features $\cuv$ (user's historical CTR/CVR on this item) represent the core information advantage of ranking over retrieval.
At retrieval time, $\cuv$ is unavailable (the target item is unknown); \DIG introduces \textbf{u2t} $\triangleq\mathbf{u2t}^{(l)}$, the token-level aggregation of $\cuv$ within bucket $s_l$, as a retrieval-time approximation.
The naming \emph{u2i} (user-to-item) and \emph{u2t} (user-to-token) reflects the coarsening from item granularity to token granularity.

\noindent\textbf{Semantic ID (SID).}
A SID is a mapping from items to discrete token sequences: $\phi:\mathcal{V}\to\mathcal{T}^L$, where $\mathcal{T}=\{1,\ldots,K\}$ is the codebook vocabulary and $L$ is the number of quantization layers.
Item $v$'s SID is denoted $\phi(v)=(s_1^v,\ldots,s_L^v)$.

\noindent\textbf{Core insight.}
Discriminative ranking finds $\arg\max_{v\in\mathcal{V}} f(u,v)$ in item space; generative retrieval finds $\arg\max_{(s_1,\ldots,s_L)} g(u,s_1,\ldots,s_L)$ in token space.
Both solve the same optimization at different granularities.
The tokenizer $\phi$ sits at the intersection: if SID construction is driven by the discriminative objective, the same token representation simultaneously carries ranking discriminability and retrieval generativity.
\emph{Generative retrieval capability is already latent inside the discriminative ranker---the tokenizer's role is to release it.}

\begin{figure*}[t]
  \centering
  \includegraphics[width=\textwidth]{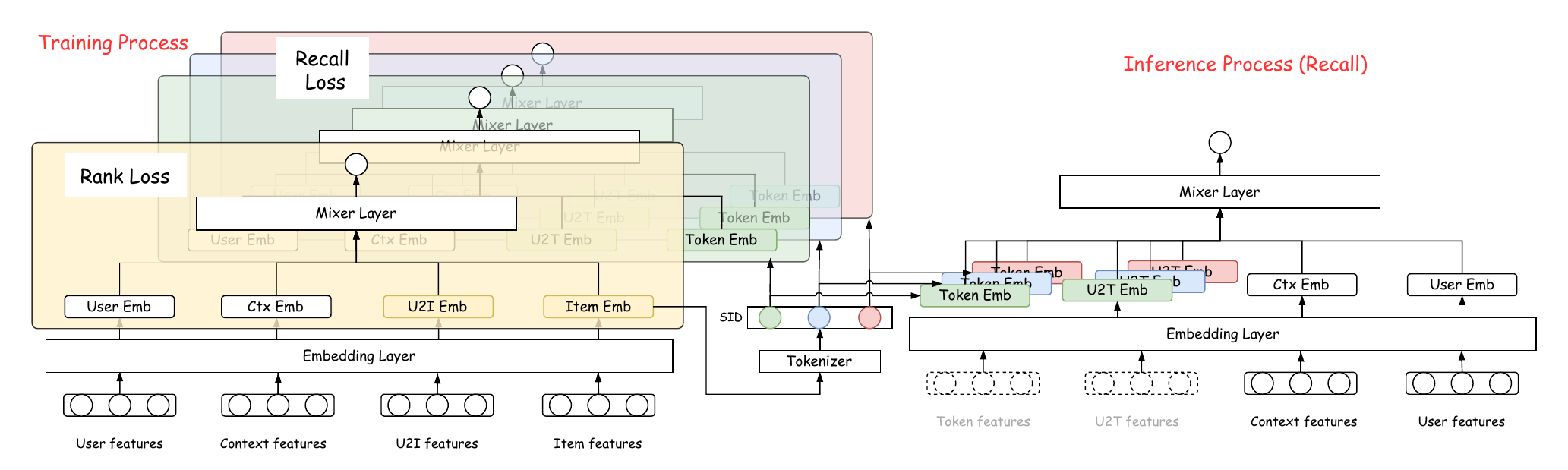}
  \caption{Overall architecture of \DIG. The tokenizer is embedded inside the DIN+DCNv2+MoE ranker. Three feature streams are handled by the feature assignment taxonomy: Type-I static features shape SIDs offline; Type-II request-level features serve as ranking/decoder conditions; Type-III u2i cross features implicitly shape codebook boundaries during training and are approximated by MLP$_\text{u2t}$ at inference.}
  \label{fig:framework}
\end{figure*}

\noindent\textbf{Feature assignment taxonomy.}
The central challenge for the unified paradigm is handling features that ranking has but retrieval does not---specifically the u2i cross features $\cuv$, which encode user-item interaction history and are the core source of discriminative ranking's personalization advantage.
These features cannot be statically encoded into SIDs (since the same item maps to different $\cuv$ for different users), yet they must somehow be preserved to avoid degrading retrieval quality.
As shown in Table~\ref{tab:feature}, \DIG partitions all features into three types, and the handling of each type is the organizing principle behind the tokenizer design (\S\ref{sec:tokenizer}), training (\S\ref{sec:training}), and inference (\S\ref{sec:inference}).

\subsection{Unified Tokenizer: SID Construction and Stability}
\label{sec:tokenizer}

\noindent\textbf{VQ encoder.}
The VQ encoder takes only item-side features $\mathbf{x}_v^s$ as input, producing a continuous item-side embedding:
\begin{equation}
\mathbf{e}_v = \mathrm{Enc}(\mathbf{x}_v^s).
\end{equation}
Restricting the encoder to time-invariant features guarantees \textbf{SID stability}: the same item always maps to the same token sequence across model versions, preserving the SID-to-item inverted index.

\noindent\textbf{Residual quantization and prefix accumulation.}
\DIG applies residual quantization (RQ) to $\mathbf{e}_v$, encoding residuals layer by layer over $L$ layers:
\begin{align}
s_l^v &= \arg\min_k \|\mathbf{r}_{l-1} - \mathbf{c}_{l,k}\|_2, \label{eq:assign}\\
\hat{\mathbf{e}}_v^{(l)} &= \mathbf{c}_{l,s_l^v}, \quad \mathbf{r}_l = \mathbf{r}_{l-1} - \hat{\mathbf{e}}_v^{(l)}, \label{eq:quantize}\\
\hat{\mathbf{e}}_v^{(1:l)} &= \textstyle\sum_{i=1}^{l} \hat{\mathbf{e}}_v^{(i)}, \label{eq:prefix}
\end{align}
where $\mathbf{r}_0 = \mathbf{e}_v$ is the initial residual; $s_l^v \in \mathcal{T}$ is the assigned code index at layer $l$; $\mathbf{c}_{l,k}$ is the $k$-th codebook vector; and $\hat{\mathbf{e}}_v^{(1:l)}$ is the prefix-accumulated quantized representation used as the item's coarse-to-fine approximation.
Codebook vectors are updated via EMA: $\mathbf{c}_{l,k}\leftarrow\alpha\,\mathbf{c}_{l,k}+(1-\alpha)\,\bar{\mathbf{e}}_{l,k}$.
The residual structure builds a coarse-to-fine hierarchy: $\hat{\mathbf{e}}_v^{(1:l)}$ approximates the full representation with monotonically shrinking error as $l$ grows---the structural basis for layer-wise beam search.

\noindent\textbf{SID–embedding decoupling.}
Existing methods use codebook vector accumulations $\hat{\mathbf{e}}_v^{(1:L)}$ directly as item representations, coupling the \emph{addressing} and \emph{semantic expression} roles of SIDs into a single set of codebook vectors.
This coupling creates a fundamental trade-off: finer SID partitions improve addressing precision but prevent semantic sharing across tokens; coarser partitions allow sharing but reduce discriminability.
Furthermore, since codebook vectors are updated via non-differentiable argmin assignment, methods that rely on them for scoring must use STE or Gumbel-Softmax to approximate gradients through the quantization step.

\DIG decouples the two roles into two separate parameter sets with different update rules:
\begin{itemize}[leftmargin=*,nosep]
  \item \textbf{Codebook vectors} $\mathbf{c}_{l,k}$: handle \emph{addressing only} (Eq.~\ref{eq:assign}--\ref{eq:quantize}), updated via EMA. Do not participate in scoring.
  \item \textbf{SID embeddings} $\mathbf{e}_{l,k}^{\mathrm{sid}}$: handle \emph{semantic expression}, updated end-to-end by the discriminative loss. Do not affect SID assignment.
\end{itemize}
The SID embedding prefix is:
\begin{equation}
\esid^{(1:l)} = \sum_{i=1}^{l} \mathbf{e}_{i,s_i^v}^{\mathrm{sid}},
\end{equation}
where $\mathbf{e}_{i,k}^{\mathrm{sid}}$ is the learnable embedding for code $k$ at layer $i$, driven end-to-end by the discriminative loss.
Because SID embeddings are standard learnable parameters independent of the argmin quantization, discriminative gradients flow through them directly---no STE approximation is needed.
This also resolves the addressing--sharing trade-off: SID partitions can be made arbitrarily fine for zero-collision addressing, while SID embeddings freely learn cross-token semantic sharing.
Critically, each SID embedding $\mathbf{e}_{l,k}^{\mathrm{sid}}$ is defined per feature field and shares the same dimensionality as the ranking item feature embeddings, so $\esid^{(1:l)}$ is a drop-in replacement for $\ev$ in the ranking MLP---the retrieval path reuses the ranking MLP without any structural modification.
Space alignment between codebook geometry and SID embeddings is maintained by $\mathcal{L}_\text{sem}$, which anchors the codebook to the encoder output by requiring the codebook vector accumulation $\hat{\mathbf{e}}_v^{(1:L)}$ to reconstruct $\mathbf{e}_v$.
Ranking and retrieval share the same $\mathrm{Mixer}$ (encompassing all modules whose parameters interact with the target item/token representation, including the scoring MLP, Target-Attention, and MoE), with $\eu$ processed through a shared BN and the remaining inputs through path-specific BNs:
\begin{align}
\hat{y}_\text{rank} &= \sigma\!\left(\mathrm{Mixer}([\mathrm{BN}_u(\eu);\,\mathrm{BN}_v(\ev);\,\mathrm{BN}_\text{u2i}(\cuv)])\right), \label{eq:rank}\\
\hat{y}_\text{recall}^{(l)} &= \sigma\!\left(\mathrm{Mixer}([\mathrm{BN}_u(\eu);\,\mathrm{BN}_\text{sid}(\esid^{(1:l)});\,\mathrm{BN}_\text{u2t}(\utot^{(1:l)})])\right), \label{eq:recall}
\end{align}
where $\mathrm{BN}_u$ is shared across both paths; $\mathrm{BN}_v$, $\mathrm{BN}_\text{u2i}$, $\mathrm{BN}_\text{sid}$, $\mathrm{BN}_\text{u2t}$ are feature-specific BNs named after the feature they normalize.

\noindent\textbf{Low-collision design.}
Since SIDs are updated end-to-end during training, strict offline zero-collision guarantees no longer apply.
\DIG instead controls collision rate through two complementary mechanisms:
(1)~\textbf{Zero-collision initialization}: codebooks are initialized via an offline Balanced K-Means tree that partitions items into equal-size clusters with zero initial collisions, providing a stable starting point for end-to-end training;
(2)~\textbf{Large SID space}: with $L=4, K=256$, the SID space supports $256^4 \approx 4$B unique tuples---orders of magnitude larger than any realistic catalog---making post-training collisions negligible in practice.

\subsection{Unified Training: Joint Loss}
\label{sec:training}

\DIG inserts a RQ quantizer after the item embedding layer of a standard DIN+DCNv2+MoE ranker; all other structures are unchanged.
The Mixer takes three inputs: item-side embedding $\ev$, user representation $\eu$, and u2i cross features $\cuv$.

\noindent\textbf{Unified training objective.}
All five losses are optimized jointly end-to-end:
\begin{equation}
\mathcal{L} = \lrank + \lrecall + \lambda_1\lcommit + \lambda_2\lsem + \lambda_3\lut.
\end{equation}

\noindent$\lrank$ is the standard ranking BCE loss that drives the whole system toward the recommendation objective:
\begin{equation}
\lrank = \mathrm{BCE}\!\left(\hat{y}_\text{rank},\,y\right).
\end{equation}

\noindent$\lrecall$ provides layer-wise supervision aligned with beam search expansion:
\begin{equation}
\lrecall = \frac{1}{L}\sum_{l=1}^{L}\mathrm{BCE}\!\left(\hat{y}_\text{recall}^{(l)},\,y\right).
\end{equation}

\noindent$\lcommit$ prevents the encoder output from drifting away from codebook entries:
\begin{equation}
\lcommit = \sum_{l=1}^{L}\|\mathrm{sg}[\mathbf{c}_{l,s_l^v}]-\mathbf{r}_{l-1}\|_2^2,
\end{equation}
where $\mathrm{sg}[\cdot]$ denotes the stop-gradient operator and $\mathbf{r}_{l-1}$ is the residual entering layer $l$.

\noindent$\lsem$ reconstructs $\mathbf{e}_v$ from the quantized representation to prevent codebook collapse:
\begin{equation}
\lsem = \|\mathbf{e}_v - \hat{\mathbf{e}}_v^{(1:L)}\|_2^2,
\end{equation}
where $\hat{\mathbf{e}}_v^{(1:L)}$ is the full-depth prefix accumulation of codebook vectors (Eq.~\ref{eq:prefix}).

\noindent$\lut$ is the MLP$_\text{u2t}$ distillation loss defined below.

\noindent\textbf{Implicit u2i shaping of the codebook (Type-III, training).}
Although $\cuv$ cannot be statically encoded into SIDs, it can indirectly drive codebook boundaries toward recommendation decision boundaries through batch aggregation.
At each training step, items in the same token bucket $s_l$ within a mini-batch $\mathcal{B}$ aggregate their u2i features into a token-level statistic:
\begin{equation}
\utot^{(l)} = \frac{1}{|\mathcal{V}_{s_l}^{(l),\mathcal{B}}|}\sum_{v'\in\mathcal{V}_{s_l}^{(l),\mathcal{B}}}\mathbf{c}_{u,v'},
\end{equation}
where $\mathcal{V}_{s_l}^{(l),\mathcal{B}}$ is the set of items in bucket $s_l$ at layer $l$ within the current mini-batch, and $\mathbf{c}_{u,v'}$ is the u2i cross feature of user $u$ for item $v'$.
Via $\lrecall$, the quantizer is implicitly rewarded for clustering items with similar u2i signals into the same bucket.
\textbf{SID boundaries thus reflect recommendation decision boundaries, not merely content similarity contours.}

\noindent\textbf{MLP$_\text{u2t}$ distillation (Type-III, training).}
Bucket-averaged $\utot^{(l)}$ loses within-bucket personalization.
MLP$_\text{u2t}$ recovers it by learning a user-conditioned approximation, supervised by the batch statistic:
\begin{align}
\utothat^{(l)}(u) &= \mathrm{MLP}_\text{u2t}^{(l)}\!\left([\eu;\,\esid^{(1:l)}]\right), \label{eq:mlpu2t}\\
\lut &= \frac{1}{L}\sum_{l=1}^{L}\|\utothat^{(l)}(u) - \mathrm{sg}[\utotbar^{(l)}(u)]\|_2^2, \label{eq:lut}
\end{align}
where $\utothat^{(l)}(u)$ is the personalized u2t prediction from $\mathrm{MLP}_\text{u2t}^{(l)}$ (one lightweight MLP per layer); $\utotbar^{(l)}(u)$ is the batch-mean u2t statistic (Eq.~14) serving as the teacher signal; and $\mathrm{sg}[\cdot]$ stops gradients from flowing into the teacher.
Inputs $\eu$ and $\esid^{(1:l)}$ are both available online at inference, so MLP$_\text{u2t}$ requires no offline feature tables.
Its output dimension equals $d_c$ (dimension of $\cuv$), so $\utothat^{(1:l)}$ can directly replace $\cuv$ in the shared MLP at retrieval time.

\subsection{Unified Inference: Feature Alignment and Beam Search}
\label{sec:inference}

\noindent\textbf{Online retrieval (beam search).}
At step $l$, all $K$ tokens at layer $l$ are scored with the shared Mixer, substituting $\cuv$ with the online MLP$_\text{u2t}$ output:
\begin{equation}
\hat{y}_\text{recall}^{(l)} = \sigma\!\left(\mathrm{Mixer}([\mathrm{BN}_u(\eu);\,\mathrm{BN}_\text{sid}(\esid^{(1:l)});\,\mathrm{BN}_\text{u2t}(\utothat^{(1:l)}(u))])\right).
\end{equation}
After $L$ steps, candidates are recovered from the SID-to-item inverted index built offline.

\noindent\textbf{Online ranking.}
Retrieved candidates enter the ranker with full u2i features:
\begin{equation}
\hat{y}_\text{rank} = \sigma\!\left(\mathrm{Mixer}([\mathrm{BN}_u(\eu);\,\mathrm{BN}_v(\ev);\,\mathrm{BN}_\text{u2i}(\cuv)])\right). \label{eq:online_rank}
\end{equation}

\noindent\textbf{Training-inference symmetry.}
Both retrieval and ranking paths share the same Mixer; the only differences are item-side representation granularity ($\esid^{(1:l)}$ at retrieval vs.\ $\ev$ at ranking) and cross-feature granularity ($\utothat^{(1:l)}$ vs.\ $\cuv$).
This symmetry fundamentally eliminates the semantic gap of traditional dual-system pipelines: the retrieval search objective is exactly aligned with the ranking optimization target.

\begin{table*}[!t]
\caption{Main retrieval results (Recall@10 / NDCG@10). Best in bold. KuaiRec datasets are retrieval-native benchmarks (full interaction matrix); Taobao and Industrial are sourced from ranking exposure logs.}
\label{tab:recall}
\centering\small
\setlength{\tabcolsep}{4.5pt}
\begin{tabular}{@{}lcccccccccc@{}}
\toprule
\multirow{2}{*}{\textbf{Method}}
  & \multicolumn{2}{c}{\textbf{Taobao}}
  & \multicolumn{2}{c}{\textbf{KuaiRec-S}}
  & \multicolumn{2}{c}{\textbf{KuaiRec-B}}
  & \multicolumn{2}{c}{\textbf{Ind-Large}}
  & \multicolumn{2}{c}{\textbf{Ind-Small}} \\
\cmidrule(lr){2-3}\cmidrule(lr){4-5}\cmidrule(lr){6-7}\cmidrule(lr){8-9}\cmidrule(lr){10-11}
  & R@10 & N@10 & R@10 & N@10 & R@10 & N@10 & R@10 & N@10 & R@10 & N@10 \\
\midrule
TIGER   & 0.0001 & 0.0020 & 0.0052 & 0.0081 & 0.0030 & 0.0045 & 0.0018 & 0.0041 & 0.0015 & 0.0033 \\
LETTER  & 0.0001 & 0.0027 & 0.0063 & 0.0109 & 0.0059 & 0.0060 & 0.0023 & 0.0051 & 0.0018 & 0.0039 \\
DAS     & 0.0003 & 0.0054 & 0.0116 & 0.0216 & 0.0102 & 0.0119 & 0.0029 & 0.0063 & 0.0021 & 0.0048 \\
DOS     & 0.0005 & 0.0106 & 0.0213 & 0.0426 & 0.0121 & 0.0234 & 0.0037 & 0.0082 & 0.0031 & 0.0067 \\
ETEGRec & 0.0007 & 0.0126 & 0.0275 & 0.0504 & 0.0153 & 0.0277 & 0.0041 & 0.0089 & 0.0035 & 0.0074 \\
\midrule
\DIG (ours) & \textbf{0.0019} & \textbf{0.0341} & \textbf{0.0418} & \textbf{0.1189} & \textbf{0.0254} & \textbf{0.0463} & \textbf{0.0097} & \textbf{0.0213} & \textbf{0.0112} & \textbf{0.0248} \\
\midrule
\textit{Improv.} & \textit{+171\%} & \textit{+171\%} & \textit{+52.0\%} & \textit{+136.1\%} & \textit{+66.0\%} & \textit{+67.1\%} & \textit{+136.6\%} & \textit{+139.3\%} & \textit{+220\%} & \textit{+235.1\%} \\
\bottomrule
\end{tabular}
\end{table*}

\section{Experiments}
\label{sec:experiments}

\subsection{Experimental Setup}
\label{sec:setup}

\noindent\textbf{Datasets.}
We evaluate on five datasets spanning extreme sparsity ranges:

\begin{itemize}[leftmargin=*,nosep]
  \item \textbf{KuaiRec-Small}~\cite{gao2022kuairec}: dense interaction matrix (1,411 users, 3,327 items, 99.6\% density). Positive label: watch\_ratio$\geq$0.7.
  \item \textbf{KuaiRec-Big}~\cite{gao2022kuairec}: sparse matrix (7,176 users, 10,728 items, 16.3\% density).
  \item \textbf{Taobao Ad}~\cite{zhu2018taobao}: real-world ad click logs ($\sim$26.6M ranking samples, $\sim$0.003\% density); u2i: per-category/brand CTR, impression counts, purchase counts.
  \item \textbf{Industrial-Large}: industrial dataset ($\sim$20M users, $\sim$500K items, $\sim$7.5M interactions, $\sim$0.0008\% density) with rich u2i cross features.
  \item \textbf{Industrial-Small}: smaller industrial business ($\sim$12K users, $\sim$40K items, $\sim$1.5M interactions, $\sim$0.31\% density, $\sim$0.016\% positive rate).
\end{itemize}

All samples are split strictly by time: each user's last click as test, second-to-last as validation, and all prior samples as training.
\emph{Ranking samples}: every exposure record with $(\mathbf{x}_u, \text{hist}, \mathbf{x}_v^s, \cuv, y)$; u2i features use prefix-accumulated statistics (no leakage).
\emph{Retrieval samples}: positive-only ($y{=}1$), full-corpus random negatives; evaluation via beam search over the full item corpus, reported as Recall@$K$ / NDCG@$K$.
KuaiRec provides a stricter \emph{retrieval-native} benchmark (full interaction matrix; test targets may never have been exposed by a ranker), while Taobao and Industrial are sourced from ranking exposure logs where the candidate pool is pre-filtered upstream.

\noindent\textbf{Backbone.}
\DIG embeds the tokenizer into a DIN+DCNv2+MoE ranker: DIN~\cite{zhou2018din} models user history via attention; DCNv2~\cite{wang2021dcnv2} captures high-order feature interactions; MoE provides multi-expert capacity for diverse user groups.
The tokenizer is a plug-in component after the item embedding layer, leaving all backbone structures intact.

\noindent\textbf{Baselines.}
We compare against five representative generative retrieval methods, all using the same NTP generative model for fair comparison:
(1)~\textbf{TIGER}~\cite{rajput2023tiger}: RQ-VAE reconstruction loss, NTP generation;
(2)~\textbf{LETTER}~\cite{li2024letter}: dual-tower contrastive learning for quantization alignment;
(3)~\textbf{DAS}~\cite{zheng2024das} and (4)~\textbf{DOS}~\cite{yin2026dos}: variants of collaborative-signal-driven quantization;
(5)~\textbf{ETEGRec}~\cite{du2024etegrec}: end-to-end tokenizer-generator joint training.

\noindent\textbf{u2i features.}
Taobao: per-category/brand CTR, impression and purchase counts (6 dims).
KuaiRec: per-user \texttt{watch\_ratio} and \texttt{like\_rate}.
Taobao/Industrial (exposure-log-based) have meaningful u2i for every training sample; KuaiRec-Big (full matrix) has 83.7\% pairs without prior interaction, making u2i features zero or degenerate for these pairs.

\noindent\textbf{Implementation.}
VQ encoder: 4-layer Transformer on item-side static features.
RQ quantizer: $L=4$ layers, $K=256$ codebook entries per layer, embedding dim $d=64$.
MLP$_\text{u2t}$: one 2-layer MLP per layer, hidden dim 64, output dim aligned with u2i cross feature dim.
Training: batch size 2048, Adam optimizer, lr $=1\times10^{-3}$, $\lambda_1=0.25$, $\lambda_2=0.1$, $\lambda_3=0.1$.
With $L=4$ layers and $K=256$ codebook entries, the u2t batch statistics are stable at all layers: layer 1 has $\sim$180 active buckets/batch ($\sim$11 items/bucket); layers 2--4 progressively narrow but the cumulative prefix $\mathbf{u2t}^{(1:l)}$ aggregates across all preceding layers, so even shallow buckets at deeper layers receive a meaningful signal.
All experiments run on 8$\times$A100 GPUs.

\subsection{Retrieval Performance}
\label{sec:recall_exp}

\noindent\textbf{RQ1: Can \DIG surpass generative retrieval baselines?}
The core claim of \DIG is that injecting discriminative signals into SID construction produces better retrieval than tokenizers trained with retrieval-only objectives.
We verify this by comparing \DIG against five representative generative retrieval baselines across all five datasets.
Table~\ref{tab:recall} reports Recall@10 and NDCG@10.

\DIG achieves the best retrieval quality across all five datasets and all metrics.
Three observations stand out:

\textbf{(1) Consistent superiority over all baselines, regardless of SID construction strategy.}
Existing tokenizers---whether trained with semantic reconstruction (TIGER) or retrieval-aligned contrastive losses (LETTER, DAS, DOS)---encode only static item attributes.
\DIG's discriminative SID aligns codebook boundaries with user preference decision boundaries via end-to-end BCE loss; beam search therefore explores a token space that reflects actual recommendation utility rather than content similarity, closing the semantic gap that plagues all NTP-based baselines.

\textbf{(2) Gains are consistent across all density regimes, with distinct underlying sources.}
On KuaiRec-Small (99.6\% density), NTP baselines already have access to strong user histories, yet \DIG still achieves +52.0\% R@10.
This gain stems from the u2t feature mechanism and end-to-end discriminative training: even in dense scenarios, u2i signals $\cuv$ drive codebook boundaries toward recommendation decision boundaries, and MLP$_\text{u2t}$ provides personalized token-level approximations that NTP models fundamentally lack.
On Taobao ($\sim$0.003\%), NTP baselines suffer additional degradation from sequence starvation (average click history length $<$5 on Taobao), causing near-complete collapse in weaker baselines (e.g., TIGER R@10$=$0.0001).
The large relative gains on Taobao (+171\% over ETEGRec) and Industrial-Small (+220\%) partly reflect baseline weakness in sparse settings; the absolute gains over the best baseline (+0.0101 and +0.0077) confirm consistent improvement.
\DIG's discriminative signal comes from ranking-side exposure samples and is independent of SID click sequence length, making it natively robust to short sequences and cold-start users.

\textbf{(3) Largest gains on industrial datasets combining sparsity with rich u2i features.}
Industrial-Small exhibits extreme interaction sparsity ($\sim$0.016\% positive rate) yet provides rich u2i cross features from the industrial ranking pipeline.
This combination delivers the highest gain across all datasets (+220\% R@10): sparsity renders NTP baselines most vulnerable (short sequences, poor generalization), while rich u2i features maximize the quality of \DIG's discriminative codebook signal---the two effects amplify each other.

\subsection{Ranking Performance}
\label{sec:rank_exp}

\noindent\textbf{RQ2: Does end-to-end training hurt ranking?}
Embedding the tokenizer inside the ranker introduces retrieval-path gradients that could interfere with the ranking objective.
We check whether this joint training hurts ranking quality by comparing \DIG against the rank-only model (recall\_loss\_weight$=$0), which has the tokenizer but disables all retrieval-path gradients.
Table~\ref{tab:ranking} reports ranking AUC across five datasets.

\begin{table}[htbp]
\caption{Ranking AUC comparison. $\Delta$ = DIG $-$ Base. Base is the rank-only model (recall\_loss\_weight$=$0).}
\label{tab:ranking}
\small
\setlength{\tabcolsep}{4pt}
\begin{tabular}{@{}lccccc@{}}
\toprule
\textbf{Method} & \textbf{Taobao} & \textbf{KuaiRec-S} & \textbf{KuaiRec-B} & \textbf{Ind-Large} & \textbf{Ind-Small} \\
\midrule
Base  & 0.6225 & 0.8428 & 0.8240 & 0.9058 & 0.6521 \\
\DIG  & \textbf{0.6402} & \textbf{0.8565} & \textbf{0.8304} & \textbf{0.9071} & \textbf{0.6726} \\
\midrule
$\Delta$ & \textbf{+0.0177} & \textbf{+0.0137} & +0.0064 & +0.0013 & \textbf{+0.0205} \\
\bottomrule
\end{tabular}
\end{table}

\textbf{End-to-end joint training consistently improves ranking quality across all five datasets.}

The primary goal of the joint training design is to not hurt ranking while enabling retrieval; the ranking results universally exceed this bar.

\noindent\textbf{Finding 1: \DIG improves ranking across all datasets.}
Gains range from +0.0013 (Industrial-Large) to +0.0205 (Industrial-Small).
On Taobao, the gain reaches +0.0177; on KuaiRec-Small, +0.0137; on KuaiRec-Big, +0.0064.
These gains come as a byproduct of the SID embedding parameter-sharing mechanism: u2i-driven codebook boundaries cluster items with similar behavior preferences, enriching SID embeddings with collaborative signals that benefit both retrieval and ranking.

\noindent\textbf{Finding 2: Larger gains on datasets with richer u2i features.}
The two largest gains occur on Industrial-Small (+0.0205) and Taobao (+0.0177), followed by KuaiRec-Small (+0.0137) and KuaiRec-Big (+0.0064)---all datasets where meaningful u2i signals are available.
This pattern confirms that the ranking bonus is not a side-effect of retrieval-path regularization, but a direct consequence of discriminative SID boundaries encoding user preference: the same codebook boundaries that drive retrieval quality also provide structured collaborative regularization to the ranking MLP.

\noindent\textbf{Finding 3: Structural mechanism behind the ranking bonus.}
Two complementary channels contribute: (1)~u2i-driven codebook boundaries provide collaborative regularization via $\lsem$ coupling between codebook geometry and SID embeddings;
(2)~layer-wise $\lrecall$ supervision encourages a hierarchical embedding structure that sharpens item discrimination at each prefix depth.
Both channels are absent in rank-only training (recall\_loss\_weight$=$0), explaining why Base consistently falls below \DIG.

\subsection{Unified Retrieval-Ranking}
\label{sec:onemodel_exp}

\noindent\textbf{RQ3: Is unified-model candidate quality better than independent retrieval?}
The unified paradigm uses the same ranking MLP for both retrieval and ranking, with MLP$_\text{u2t}$ approximating u2i features at token granularity during retrieval.
The key question is whether this approximation preserves enough scoring quality to make the retrieval path useful---and whether the resulting retrieval-ranking AUC gap is acceptable.
Table~\ref{tab:onemodel} quantifies this gap across datasets.

\begin{table}[htbp]
\caption{Unified retrieval-ranking AUC gap. The gap measures how much the retrieval path's scoring ability lags behind the full ranking path.}
\label{tab:onemodel}
\small
\setlength{\tabcolsep}{5pt}
\begin{tabular}{@{}lcccc@{}}
\toprule
\textbf{Dataset} & \textbf{Rank AUC} & \textbf{Recall AUC} & \textbf{Gap ($\Delta$)} \\
\midrule
Taobao      & 0.6402 & 0.6115 & $-$0.0287 \\
KuaiRec-S   & 0.8565 & 0.8431 & $-$0.0134 \\
KuaiRec-B   & 0.8304 & 0.8132 & $-$0.0172 \\
Ind-Large   & 0.9071 & 0.8459 & $-$0.0612 \\
Ind-Small   & 0.6726 & 0.6483 & $-$0.0243 \\
\bottomrule
\end{tabular}
\end{table}

\noindent\textbf{Finding 1: The retrieval-ranking AUC gap correlates with u2i feature complexity, not interaction density.}
On KuaiRec datasets, u2i features are simple (watch\_ratio, like\_rate), and MLP$_\text{u2t}$ approximates them accurately---gaps are small ($-$0.0134 on KuaiRec-S, $-$0.0172 on KuaiRec-B).
On Taobao and Industrial-Large, u2i features are high-dimensional (per-category/brand CTR, impression counts, purchase counts), making the token-level approximation harder and widening the gap ($-$0.0287 and $-$0.0612 respectively).
This gap is an inherent cost of the unified-model design: MLP$_\text{u2t}$ approximates item-level $\cuv$ at token granularity, and approximation error grows with u2i feature complexity.

\subsection{Ablation Study}
\label{sec:ablation}

\DIG's design involves three key components: (1)~end-to-end discriminative gradients reshaping the codebook, (2)~training-side u2i signals implicitly shaping bucket boundaries, and (3)~MLP$_\text{u2t}$ providing personalized u2t at inference.
We isolate each component to quantify its individual contribution and verify the necessity of the full design.

\noindent\textbf{Ablation 1: Discriminative gradient necessity (E2E vs. two-stage).}
We replace \DIG's end-to-end training with a two-stage baseline (Fixed SID): SIDs are pre-generated offline using the same Balanced K-Means tree initialization as \DIG, then fixed throughout training---the VQ encoder is frozen and codebook boundaries do not change.
This isolates the sole variable of whether discriminative gradients are allowed to reshape the codebook, evaluated across all three public datasets.

\begin{table}[htbp]
\caption{Ablation 1: Effect of discriminative gradient on retrieval and ranking across three public datasets.}
\label{tab:abl1}
\small
\setlength{\tabcolsep}{4pt}
\begin{tabular}{@{}llccc@{}}
\toprule
\textbf{Dataset} & \textbf{Config} & \textbf{Rank AUC} & \textbf{Recall AUC} & \textbf{Recall@10} \\
\midrule
\multirow{3}{*}{Taobao}
  & \DIG (E2E)  & \textbf{0.6402} & \textbf{0.6115} & \textbf{0.0019} \\
  & Fixed SID   & 0.6355          & 0.6089          & 0.0014 \\
  & $\Delta$    & $+$0.0047       & $+$0.0026       & $+$35.7\% \\
\midrule
\multirow{3}{*}{KuaiRec-S}
  & \DIG (E2E)  & \textbf{0.8565} & \textbf{0.8431} & \textbf{0.0418} \\
  & Fixed SID   & 0.8544          & 0.7562          & 0.0275 \\
  & $\Delta$    & $+$0.0021       & \textbf{+0.0869}& \textbf{+52.0\%} \\
\midrule
\multirow{3}{*}{KuaiRec-B}
  & \DIG (E2E)  & \textbf{0.8304} & \textbf{0.8132} & \textbf{0.0254} \\
  & Fixed SID   & 0.8308          & 0.7451          & 0.0008 \\
  & $\Delta$    & $-$0.0004       & \textbf{+0.0681}& \textbf{+212\%} \\
\bottomrule
\end{tabular}
\end{table}

Across all three datasets, fixing the tokenizer consistently degrades Recall AUC (KuaiRec-S: $-$0.0869; KuaiRec-B: $-$0.0681; Taobao: $-$0.0026), while Rank AUC remains virtually identical ($|\Delta|\!\leq\!0.002$).
The Recall AUC gaps are especially pronounced on KuaiRec datasets, where the fixed tokenizer groups items by content geometry rather than user preference---beam search explores semantically coherent but preference-irrelevant regions.
The near-zero ranking cost across all datasets confirms that discriminative gradients reshape the codebook without interfering with the ranking MLP, which still receives full u2i features regardless.
\textbf{Discriminative gradient is the fundamental source of \DIG's retrieval quality, with negligible cost to ranking.}

\noindent\textbf{Ablation 2: Independent contributions of training-side u2i and inference-side MLP$_\text{u2t}$.}
Orthogonal ablation on Taobao:

\begin{table}[htbp]
\caption{Ablation 2: Orthogonal ablation on Taobao (R@10 / N@10).}
\label{tab:abl2}
\small
\begin{tabular}{@{}lcc@{}}
\toprule
\textbf{Config} & \textbf{R@10} & \textbf{N@10} \\
\midrule
Full \DIG               & 0.0019 & 0.0341 \\
w/o training-side u2i   & 0.0014 ($-$26.3\%) & 0.0253 ($-$25.8\%) \\
w/o inference MLP$_\text{u2t}$ & 0.0016 ($-$15.8\%) & 0.0290 ($-$14.9\%) \\
w/o both                & 0.0010 ($-$47.4\%) & 0.0183 ($-$46.3\%) \\
\bottomrule
\end{tabular}
\end{table}

The complementary effect confirms mutual reinforcement: removing both causes $-$47.4\%, exceeding the sum of individual drops ($-$26.3\%$+$$-$15.8\%$=$42.1\%).
Training-side u2i drives codebook boundaries toward recommendation decision boundaries, which in turn improves MLP$_\text{u2t}$ distillation signal quality---the two components amplify each other rather than independently stacking.

\noindent\textbf{Ablation 3: MLP$_\text{u2t}$ vs. statistical mean u2t.}
On Taobao: MLP$_\text{u2t}$ (full \DIG) achieves R@10=0.0019, while statistical mean u2t yields R@10=0.0015 ($-$21.1\%).
MLP$_\text{u2t}$'s advantage is most prominent on \emph{sparse tokens and long-tail users}: for low-activity users, statistical mean u2t is highly unstable, while MLP$_\text{u2t}$ recovers stable estimates through model generalization.

\noindent\textbf{Ablation 4: Necessity of layer-wise supervision.}
Replacing layer-wise $\lrecall$ with only final-layer supervision ($l=L$) causes retrieval metrics to drop significantly ($-$18.4\% R@10) while ranking AUC remains nearly unchanged ($\approx 0$).
This precisely separates two mechanisms: layer-wise supervision's value is \emph{eliminating training-inference discrepancy} (intermediate prefixes become out-of-distribution inputs without supervision), not providing ranking regularization.
DIG's positive ranking effect comes from structural improvement of discriminative SID embeddings, independent of layer-wise supervision.

\subsection{Sparse Scenario Stability Analysis}
\label{sec:sparse_exp}

\DIG relies on u2i signals to drive codebook boundaries toward recommendation decision boundaries (via $\lrecall$) and to supervise MLP$_\text{u2t}$ distillation (via batch u2t statistics).
A natural concern is whether \DIG remains stable when u2i features are sparse or degenerate---a common situation in real-world deployments where not all user-item pairs have interaction history.
We systematically analyze stability across five datasets with varying u2i completeness (Table~\ref{tab:sparse}).

\begin{table}[htbp]
\caption{Sparse scenario stability. $\Delta$AUC = \DIG $-$ Base ranking AUC.}
\label{tab:sparse}
\small
\setlength{\tabcolsep}{3pt}
\begin{tabular}{@{}lcccc@{}}
\toprule
\textbf{Dataset} & \textbf{Density} & \textbf{u2i Completeness} & \textbf{$\Delta$AUC} & \textbf{Stability} \\
\midrule
KuaiRec-S  & 99.6\% & Full & +0.0137 & Stable \\
KuaiRec-B  & 16.3\% & Sparse & +0.0064 & Stable \\
Taobao     & 0.003\% & Full (ranking logs) & +0.0177 & Stable \\
Ind-Large  & 0.0008\% & Full$^*$ & +0.0013 & Stable \\
Ind-Small  & 0.31\% & Full$^*$ & +0.0205 & Stable \\
\bottomrule
\end{tabular}

{\footnotesize $^*$ Industrial datasets use hundreds of u2i cross features sourced directly from real industrial ranking pipelines, providing full coverage over all exposed user-item pairs.}
\end{table}

\noindent\textbf{Finding 1: Training sample construction determines u2i completeness, not interaction matrix density.}
Taobao's interaction matrix density is only 0.003\%, yet every training sample carries complete u2i features because the data comes from ranking exposure logs---the ranker only scores user-item pairs that already have interaction history.
KuaiRec-Big has 16.3\% matrix density, but its training set is built from the full interaction matrix including unobserved pairs---83.7\% of training samples have zero or degenerate u2i values.
Despite this, \DIG still achieves +0.0064 AUC gain on KuaiRec-Big, suggesting that even partial u2i coverage is sufficient for discriminative SID boundaries to provide ranking benefit.

\noindent\textbf{Finding 2: $\lrecall$ is the primary training signal for SID embeddings.}
The reason is structural: the ranking path uses full item-side embedding $\ev$, so $\lrank$'s gradient flows mainly through $\ev$ and $\eu$---SID embeddings receive only a weak indirect signal via $\lsem$.
$\lrecall$ directly supervises the SID prefix $\esid^{(1:l)}$ at every layer, making SID embeddings its primary optimization target.
Setting recall\_loss\_weight to 0 causes SID embeddings to degrade and ranking AUC to fall, confirming that $\lrecall$ is irreplaceable.

\noindent\textbf{Conclusion.}
\DIG is stable across all five datasets regardless of interaction matrix density.
The key driver is the quality of u2i signals in ranking exposure logs; \DIG leverages these signals to reshape codebook boundaries, delivering consistent ranking gains even in sparse interaction scenarios.
For industrial deployment, \textbf{u2i feature coverage rate should be evaluated first; for low-coverage scenarios, configuring a stronger MLP$_\text{u2t}$ further improves distillation quality}.

\section{Conclusion}
\label{sec:conclusion}

We proposed \DIG, a unified paradigm that bridges discriminative ranking and generative retrieval by embedding the tokenizer inside the ranker for end-to-end joint training.
The key insight---ranking and retrieval are the same optimization at different granularities---motivates a feature assignment taxonomy that governs all three design components: the tokenizer decouples SID addressing from semantic expression via independent SID embeddings; a five-part unified loss drives both paths jointly, with u2i signals driving codebook boundaries toward recommendation decision boundaries and MLP$_\text{u2t}$ bridging the feature gap at inference; and the shared ranking MLP executes beam search directly, eliminating the semantic gap of traditional dual-system pipelines.
Experiments on five datasets confirm that \DIG simultaneously improves retrieval, ranking, and unified-model candidate quality, with especially strong gains in sparse and u2i-rich scenarios.

\bibliographystyle{ACM-Reference-Format}
\bibliography{references}

\end{document}